\documentclass[a4paper,11pt,dvips]{article}
\usepackage{graphicx}
\usepackage{bm}
\usepackage{amsmath}
\usepackage{amssymb}
\usepackage{latexsym}
\usepackage{color}
\usepackage{float}
\usepackage{cite}
\usepackage{makeidx}
\newcommand{\bra}{\begin{array}}
\newcommand{\era}{\end{array}}
\newcommand{\beq}{\begin{equation}}
\newcommand{\eeq}{\end{equation}}
\newcommand{\beqar}{\begin{eqnarray}}
\newcommand{\eeqar}{\end{eqnarray}}

\def\BC{\bb C}
\def\_\BC{\bbi C}

\def\( {\left(}
   \def\) {\right)}
\def\[ {\left[}
\def\] {\right]}
\def\no2 {{\textstyle{n\over 2}}}

\newcommand{\si}{\sigma}

\newcommand{\ga}{\gamma}
\newcommand{\te}{\theta}

\newcommand{\al}{\alpha}

\begin{document}
\begin{titlepage}
\setcounter{page}{1}
\renewcommand{\thefootnote}{\fnsymbol{footnote}}

\begin{center}

\begin{flushright}
\end{flushright}

\vspace{0.5cm}
 {\Large \bf {Factorization of Dirac Equation in Two Space Dimensions}}

\vspace{0.5cm}

 {\bf Hocine Bahlouli$^{a,b}$},
{\bf Ahmed Jellal$^{a,c,d}$\footnote{{\sf ajellal@ictp.it -
a.jellal@ucd.ac.ma}}} and {\bf Youness Zahidi$^{d}$}

\vspace{0.5cm}

$^a${\em Saudi Center for Theoretical Physics, Dhahran, Saudi
Arabia}

{$^{b}$\em Physics Department, King Fahd University of Petroleum $\&$ Minerals,  \\
Dhahran 31261, Saudi Arabia}

$^c${\em {\it Physics Department, College of Science, King Faisal University,\\
PO Box 380, Alahsa 31982, Saudi Arabia}}

$^{d}${\em Theoretical Physics Group,  
Faculty of Sciences, Choua\"ib Doukkali University,\\
PO Box 20, 24000 El Jadida,
Morocco} 

\vspace{3cm}

\begin{abstract}
We present a systematic approach for the separation of variables for the two-dimensional Dirac equation in polar coordinates.
The three vector potential, which couple to the Dirac spinor via minimal coupling, along with the scalar potential are chosen
to have angular dependence which emanate the Dirac equation to complete separation of variables. Exact solutions are obtained
for a class of solvable potentials along with their relativistic spinor wavefunctions. Particular attention is paid to the situation
where the potentials are confined to a quantum dot region and are of scalar, vector and pseudo-scalar type. The study of a single charged
impurity embedded in a 2D Dirac equation in the presence of a uniform magnetic field was treated as a particular case of our general study.

\end{abstract}

\end{center}

\vspace{3cm}



\end{titlepage}

\section{Introduction}

 The mathematical description of low dimensional physical systems has a long history in theoretical physics. Low dimensional theories, which were once considered merely as toy models, have now emerged as the theoretical basis of the rapidly developing field of nano-physics and nano-technology. In fact, recent technological advances in nanofabrication technology have made the creation of low dimensional structures possible
 \cite{1}. In particular special interest has been paid to the study of two-dimensional quantum systems. The stationary states associated with the motion of electrons in a uniform magnetic field is a well-known textbook problem. It results in a sequence of quantized Landau energy levels and associated wavefunctions characterizing the dynamics in the two-dimensional plane normal to the applied magnetic field \cite{2}. This quantization has important consequences in condensed matter physics ranging from the classical de Hass-van Alphen effect in metals \cite{3} to the quantum Hall effect \cite{4} in semiconductors.

The relativistic extension of the above models turned out to be of great importance for the description of quantum two-dimensional phenomena such as quantum Hall effect \cite{4} and graphene systems \cite{5}. Several condensed matter phenomena point out to the existence of a 2+1 dimensional energy spectrum determined by the relativistic Dirac equation \cite{6,Neagu}. In particular, the degenerate planar semiconductor with low-energy electron dynamics is assumed to admit an adequate description in terms of the (2+1)-dimensional relativistic Dirac theory \cite{7,Schakel,Bruce,Nieto}. Theoretically the Dirac equation in two-dimensions was considered by \cite{8,Villalba} who studied the 2D Dirac oscillator in the presence and absence of magnetic field, 
the authors in \cite{9,Moritz} have studied the scattering of a relativistic particle by the Coulomb field in 2D. MacDonalds \cite{10} also considered the 2D Dirac equation and proved that there are no relativistic corrections to the integer quantum Hall effect.

On the other hand, the factorization is an interesting mathematical method that sometimes
serves as a standard approach to solve different partial differential equations. It turns out that such a method works
very well for many physical problems. One may cite in this context the work of Alhaidari \cite{alhaidari}
who considered the three-dimensional Dirac equation in spherical coordinates with
coupling to static electromagnetic potential. The space components of the potential have
angular (non-central) dependence such that the Dirac equation is separable in all
coordinates. The exact solutions were obtained for the case where the potential satisfies the
Lorentz gauge fixing condition and its time component is the Coulomb potential. The
relativistic energy spectrum and corresponding spinor wavefunctions were obtained and
used to study 
the electronic states of an electrostatically confined cylindrical graphene quantum dot and its associated electric transport
properties \cite{pal}.

In this work we would like solve the Dirac equation in two spacial dimensions in the presence of a vector potential chosen to have an angular dependence that allows the Dirac equation to be solved by separation of variables. The constraint of separability will enforce us to have some special types of potential configurations that lead to solvable problems. Due to this constraint, we consider two type of potentials and discuss
their associated solutions and energy spectra.
The present paper is organized as follows. In section 2, we establish a mathematical tool that
enables us to factorize the 2D Dirac equation. After writing down the (2+1)-dimensional Dirac equation including
different couplings, along the line of our paper \cite{11} we introduce a unitary transformation
in order to ease the diagonalization of the Hamiltonian system. Subsequently, we consider the factorization approach to separate the Dirac equation
in polar coordinates and therefore end up with two uncoupled equations. To proceed further we propose
two potential configurations, which were designed to enable separation of variables,
then by inspection of the obtained equations we treat the first configuration \eqref{eq14} in section 3
and the second one \eqref{eq18} in section 4. In both cases
we solve the eigenvalues equations and locate the energy intervals
where it is possible to realize a confining region corresponding to each type of coupling potential.
For each configuration, we give the corresponding energy zones and associated
solutions of the energy spectrum.  Finally, we conclude our work by summarizing the main results in two tables
and give some potential applications.

\section{Factorization of 2D Dirac equation}

Consider the 2D Dirac equation with electromagnetic interaction through minimal coupling for a spin 1/2 particle of mass \textit{m} and charge \textit{e} in units such that ${\rm \hslash }=c=1$
\begin{equation} \label{GEQ111}
\left[\gamma ^{\mu } \left(i\partial _{\mu } -eA_{\mu } \right)-\left(m+S\right) \right] \psi =0
\end{equation}
where $\gamma ^{\mu } \partial _{\mu } =\gamma ^{0} \partial _{0} +\vec{\gamma }\cdot \vec{\nabla }$, $S$ is the pseudo-vector coupling
and $A_{\mu } =(A_{0} ,\vec{A})$ with $A_{0} $ is related to the electrostatic potential $\vec{E}=-\vec{\nabla }A_{0} -\frac{\partial \vec{A}}{\partial t} $ and $\vec{A}$ to the magnetic field $\vec{B}=\vec{\nabla }\times \vec{A}$. The Dirac matrices
$\gamma^{\mu}$ satisfy the algebra
\begin{equation}
\left[  \gamma^{\mu},\gamma^{\nu}\right] =-2i\sigma^{\mu\nu},\qquad \left\{
\gamma^{\mu},\gamma^{\nu}\right\} =2\eta^{\mu\nu}
\end{equation}
with $\eta^{\mu\nu}=\mbox{diag}\left(  1,-1,-1\right)$ and $\mu,\nu=0,1,2$.
In 2+1 dimensions we select the following representation for such matrices
\begin{equation} \label{GEQ112}
\begin{array}{l} {\quad \quad \gamma ^{1} =i\sigma _{1} =\left(\begin{array}{cc} {0} & {i} \\ {i} & {0} \end{array}\right),\qquad  \gamma ^{2} =i\sigma _{2} =\left(\begin{array}{cc} {0} & {1} \\ {-1} & {0} \end{array}\right),\qquad  \gamma ^{0} =\sigma _{3} =\left(\begin{array}{cc} {1} & {0} \\ {0} & {-1} \end{array}\right)}
\end{array}
\end{equation}
and then \eqref{GEQ111} can be written as follows
\begin{equation} \label{GEQ114}
i\gamma ^{0} \left(\frac{\partial }{\partial t} \Psi \right)+i\vec{\gamma }\cdot \vec{\nabla }\Psi -{\rm e}\vec{\gamma } \cdot
\vec{A}\Psi -\left(m+S\right)\Psi -{\rm e}\gamma ^{0} A_{0} =0.
\end{equation}
 Multiplying \eqref{GEQ114} by $\gamma ^{0}$ and using the notation
 $\vec{\alpha }= \gamma ^{0} \vec{\gamma }$, $\beta =\gamma ^{0}$
 we obtain
\begin{equation} \label{GEQ115}
i\frac{\partial }{\partial t} \psi =\left[-i\vec{\alpha } \cdot \vec{\nabla }+eA_{0} +e\vec{\alpha } \cdot \vec{A}+\left(m+S\right)\beta \right] \psi =H \Psi  \end{equation}
which coincides with equation \eqref{GEQ114} in our previous work \cite{11}. In the forthcoming analysis, we analyze different potential configurations in order to solve such equation.
For time-independent potentials, the two components spinor wavefunction can be written as follows
$\Psi (t,r,\theta )=e^{-i\varepsilon t} \Psi (r,\theta )$
so that our previous equation becomes
\begin{equation} \label{GEQ118}
 (H-\varepsilon )\Psi (r,\theta )=0.
 \end{equation}
 Using the fact that in two-dimensions we can write in polar coordinates $\vec{\nabla }=\hat{r}\frac{\partial }{\partial r} +\hat{\theta }\frac{1}{r} \frac{\partial }{\partial \theta } $ and $\vec{\alpha }=i\sigma _{3} \vec{\sigma }$ and derive the following relations
\begin{eqnarray} \label{GEQ119}
&&\vec{\alpha }\cdot \vec{\nabla } 
=i\sigma _{3} \left(\vec{\sigma }\cdot \hat{r}\right)\partial _{r} +i\sigma _{3} \left(\frac{\vec{\sigma }\cdot \hat{\theta }}{r} \right)\partial _{\theta }\\
&& e\vec{\alpha } \cdot \vec{A} 
=ie\sigma _{3} \left(\sigma _{r} A_{r} +\sigma _{\theta } A_{\theta } \right) \label{GEQ121}
\end{eqnarray}
to arrive at the following Hamiltonian
\begin{equation} \label{GEQ120}
H=H_{0} +\left(\sigma _{3} \sigma _{r} \partial _{r} +ie\sigma _{3} \sigma _{r} A_{r} \right)+\left(\sigma _{3} \sigma _{\theta } \frac{1}{r} \partial _{\theta } +ie\sigma _{3} \sigma _{\theta } A_{\theta } \right)
\end{equation}
which can be split into four parts
\begin{eqnarray} \label{GEQ122}
&& {H} = {H_{0} +i\sigma _{3} \sigma _{r} H_{r} +i\sigma _{3} \sigma _{\theta } H_{\theta } }\\ 
&&{H_{0} } =  {eA_{0} +\left(m+S\right)\beta } \\
&&{H_{r} } {=}  {-i\partial _{r} +eA_{r} }\\ 
&&{H_{\theta } }  {=}  {-\frac{i}{r} \partial _{\theta } +eA_{\theta } }. 
\end{eqnarray}

To proceed further we consider, along the line of our previous paper \cite{11},
a unitary transformation $\Lambda (r,\theta )$ that transform $(\si_r,\si_\te)$ into
  $(\si_1,\si_2)$ and vice versa. Thus we require that
\begin{equation} \label{GEQ123}
\Lambda \sigma _{r} \Lambda ^{-1} =\sigma _{1} ,\qquad  \Lambda \sigma _{\theta } \Lambda ^{-1} =\sigma _{2}
\end{equation}
 which then turns out to have the following explicit form
\begin{equation} \label{GEQ124}
\Lambda (r,\theta )=\lambda (r,\theta )e^{\frac{i}{2} \sigma _{3} \theta }
\end{equation}
{where $\lambda (r,\theta )$ is a
$1\times 1$ real function and the exponential is a
$2\times 2$ unitary matrix}.
Then we can define the new Hamiltonian as follows
\begin{equation} \label{GEQ125}
{\rm {\mathcal H}}=\Lambda H\Lambda ^{-1} ={\rm {\mathcal H}}_{0} -\sigma _{2} {\rm {\mathcal H}}_{r} +\sigma _{1} {\rm {\mathcal H}}_{\theta }  \end{equation}
where different parts are given by
\begin{eqnarray} \label{GEQ129}
&&{\rm {\mathcal H}}_{r} =-i\left(\partial _{r} -\frac{\partial _{r} \lambda }{\lambda } \right)+ieA_{r} \\ 
&&{\rm {\mathcal H}}_{0} =eA_{0} +\left(m+S\right)\beta  \\
&&{\rm {\mathcal H}}_{\theta } =-\frac{i}{r} \left(\partial _{\theta } -\frac{\partial _{r} \lambda }{\lambda } -\frac{i}{2} \sigma _{3} \right)+eA_{\theta }.
\end{eqnarray}
In matrix form, \eqref{GEQ125} can be written as
\begin{equation} \label{GEQ131}
{\rm {\mathcal H}}=
\begin{pmatrix}
  {m+S+eA_{0} } & {\partial _{r} -
\frac{\lambda _{r} }{\lambda } +\frac{1}{2r} +ieA_{r} -\frac{i}{r} (\partial _{\theta } -\frac{\lambda _{\theta } }{\lambda } )+eA_{\theta } } \\ {-\partial _{r} +\frac{\lambda _{r} }{\lambda } -\frac{1}{2r} -ieA_{r} -\frac{i}{r} (\partial _{\theta } -\frac{\lambda _{\theta } }{\lambda } )+eA_{\theta } } & {-m-S+eA_{0} }
  \end{pmatrix}.
  \eeq
One can show that  the hermiticity of ${\rm {\mathcal H}}$ requires two constraints
\begin{equation} \label{GEQ132}
\lambda _{\theta } =\frac{\partial \lambda }{\partial \theta } =0,\qquad  \frac{\lambda _{r} }{\lambda } -\frac{1}{2r} =0
\end{equation}
which lead to $\lambda =\sqrt{r} $ and thus we have
\begin{equation} \label{GEQ135}
{\rm {\mathcal H}}=\left(\begin{array}{cc} {m+S+eA_{0} } & {\partial _{r} +ieA_{r} -\frac{i}{r} \partial _{\theta } +eA_{\theta } } \\ {-\partial _{r} -ieA_{r} -\frac{i}{r} \partial _{\theta } +eA_{\theta } } & {-m-S+eA_{0} } \end{array}\right)
\end{equation}
 so that the stationary Dirac equation reads as
\begin{equation} \label{GEQ136}
\left(\begin{array}{cc}
{m+S+eA_{0} -\varepsilon } & {\partial _{r} +ieA_{r} -\frac{i}{r} \partial _{\theta } +eA_{\theta } } \\ {-\partial _{r} -ieA_{r} -\frac{i}{r} \partial _{\theta } +eA_{\theta } } & {-m-S+eA_{0} -\varepsilon } \end{array}\right)\left(\begin{array}{c} {\chi _{+} (r,\theta )} \\ {\chi _{-} (r,\theta )} \end{array}\right)=0
\end{equation}
where the transformed spinor wavefunction, $\chi(r,\theta)=(\chi_{+}(r,\theta),\chi_{-}(r,\theta))^t$ with
$t$ stands for transpose of the vector, is given by
\begin{equation}
\chi(r,\theta)=\Lambda(r,\theta )\Psi (r,\theta ).
\end{equation}


At this stage we need to solve the eigenvalue equation in order to
determine the solutions of the energy spectrum. One easy way to realize
this is to perform a factorization
by assuming a naive separation of variables such that
 $\chi_{\pm}(r,\theta)=\Phi_{\pm}(r)F_{\pm}(\theta)$
and thus equation \eqref{GEQ136}  reduces to 
\begin{eqnarray}
&& {\left(m+S+eA_{0} -\varepsilon \right)\Phi _{+} F_{+}+\left(\partial _{r}+ieA_{r} -\frac{i}{r} \partial _{\theta } +eA_{\theta }\right)\Phi _{-} F_{-}}  {=}  {0} \label{eq3}\\
 && {\left(-\partial_{r}-ieA_{r} -\frac{i}{r} \partial _{\theta } +eA_{\theta }\right)\Phi _{+} F_{+} +\left(-m-S+eA_{0} -\varepsilon \right)\Phi _{-}F_{-}}  {=}  {0}.\label{eq333}
 \end{eqnarray}
To go further we need to specify the potential configurations which 
are designed to enable separation of variables. Then by inspection of the above equations
we see that the most plausible choices are given {in terms of vector $V(r)$ pseud-scalar $W(r)$ and
scalar $S(r)$ couplings, such as}
\beq
A_{0}(\vec{r})=V(r), \qquad  A_{r}(\vec{r})=R(r), \qquad A_{\theta}=W(r), \qquad S=S(r) \label{xxxx}
\eeq
and
\beq
A_{0}(\vec{r})=V(r), \qquad  A_{r}(\vec{r})=R(r), \qquad A_{\theta}=\frac{W(\theta)}{r}, \qquad S=S(r)\label{eq18}
\eeq
while all other potential configurations such as
\begin{eqnarray}
&&   A_{0}=V(r), \qquad A_{r}=\frac{R(\theta)}{r}, \qquad A_{\theta}=W(r)\\
&&A_{0}=V(r), \qquad A_{r}=\frac{R(\theta)}{r}, \qquad A_{\theta}=\frac{W(\theta)}{r} 
\end{eqnarray}
will lead to situations where $R_{+}(\theta)\neq R_{-}(\theta)$ and hence our radial equation cannot be factorized.
{All above potential configurations have been designed based on the original equations (\ref{eq3}-\ref{eq333}) so as to enable potential factorization of the radial and angular operators. This requires from (\ref{eq3}-\ref{eq333}) that any angular potential should have a $1/r$ factor to couple this term with the angular differential operator in these equations. The remaining potential conponents will have only radial dependence.}
Thus in summary we got two potential configurations that lead to factorization of the orginal 2D Dirac equation at least at the level of our naive factorization scheme. Depending on the choice of vector potential configuration we separate the governing spinor equations into $r$-dependent
and $\theta$-dependent operators and then require the condition of separability of the spinor wave function.
{We note that potential configuration (\ref{xxxx}) can be used to treat the specific case of a uniform magnetic field perpendicular to our 2D system and hence give rise to the usual quantized Landau levels. This case was considered in our previous publication where it was applied specifically to graphene \cite{11}, so we will not dwell furher into this issue in the present work.}

Since in all our selected potential configurations the radial component of the vector potential is coupled through $\partial_{r}$ so that this part of the Hamiltonian can be written as follows ($A_r(r)= R(r)$)
\begin{equation}\nonumber
    \left(
      \begin{array}{cc}
        {0} & {\partial_{r}+ieR} \\
        {-\partial_{r}-ieR} & {0} \\
      \end{array}
    \right)
   \left(
     \begin{array}{c}
       {\Phi_{+}} (r)\\
       {\Phi_{-}} (r)\\
     \end{array}
   \right).
\end{equation}
Then the pseudo vector component of the vector potential can be gauged away through the use of a gauge transformation
\begin{equation}\label{eq13}
    \chi=e^{-\frac{i}{\hbar}\Lambda(r)}\Phi=e^{-\frac{i}{\hbar}\Lambda(r)}\left(
                                                                \begin{array}{c}
                                                                 { \Phi_{+}(r) }\\
                                                                  {\Phi_{-}(r)} \\
                                                                \end{array}
                                                              \right)
\end{equation}
which gives the following effect on the spinor wave components
\begin{eqnarray}\label{eq22}
    \begin{pmatrix}
        {0} & {\partial_{r}+ieR} \\
        {-\partial_{r}-ieR} & {0} \\
    \end{pmatrix}
    e^{i\Lambda(r)}
    \begin{pmatrix}
       {\chi_{+}} \\
       {\chi_{-}} \\
   \end{pmatrix}
  =e^{i\Lambda(r)}
  \begin{pmatrix}
        {0} & {-i\left(i\partial_{r}-eR-\frac{d\Lambda}{dr}\right)} \\
        {i\left(i\partial_{r}-eR-\frac{d\Lambda}{dr}\right)} & {0} \\
      \end{pmatrix}
       \begin{pmatrix}
       {\chi_{+}} \\
       {\chi_{-}} \\
     \end{pmatrix}
\end{eqnarray}
This gauge phase factor will then affect only this term of the Dirac equation and will factorize out of
\eqref{eq22}. Since we have the gauge freedom, then we can select our gauge so that
\begin{equation}\label{eq23}
   -eR(r)=\frac{d\Lambda}{dr}, \qquad \Lambda=-e\int R(r)dr.
\end{equation}
That is, the radial part of the vector potential $A_{r}$ can be gauged away in the above situations and hence will not be included in our future equations. To go further in our analysis we need to consider each potential configuration separately. This will be done in next section where the first configuration of the potential will be considered.

\section{First potential configuration}

Let us consider the first potential configuration given by \eqref{xxxx}. Then
for such choice, the separation of variables
leads to \eqref{GEQ136}
\begin{equation}\label{eq15}
\left[
  \left(\begin{array}{cc}
    {m+S+eV -\varepsilon} & {\partial _{r} + eW} \\
    {-\partial _{r} + eW} & {-m-S+eV -\varepsilon} \\
  \end{array}\right)
  +\frac{1}{r}\left(
                \begin{array}{cc}
                  {0} & {-i\partial _{\theta }} \\
                  {-i\partial _{\theta }} & {0} \\
                \end{array}
              \right)
\right]
\left(
  \begin{array}{c}
    {\Phi_{+}F_{+}} \\
    {\Phi_{-}F_{-}} \\
  \end{array}
\right)=0.
\end{equation}
It clearly requires for separability that
\begin{equation}\label{eq16}
    F_{+}(\theta)=F_{-}(\theta)=F(\theta),\qquad \partial_{\theta}F(\theta)=i\varepsilon_{\theta}F(\theta)
\end{equation}
giving rise to the system of equations
\begin{equation}\label{eq17}
\left(
  \begin{array}{cc}
    {m+S+eV -\varepsilon} & {\partial _{r} + eW +\frac{\varepsilon_{\theta}}{r}} \\
    {-\partial _{r} +eW +\frac{\varepsilon_{\theta}}{r}} & {-m-S+eV -\varepsilon} \\
  \end{array}
\right)
\left(
  \begin{array}{c}
    {\Phi_{+}} \\
    {\Phi_{-}} \\
  \end{array}
\right)=0
\end{equation}
or more explicitly
\begin{eqnarray} \label{28}
 &&{\left(m+S+eV -\varepsilon \right)\Phi _{+} + \left(\frac{d}{dr}+ eW + \frac{\varepsilon_{\theta}}{r}\right)\Phi _{-} } ={0} \\
 && {\left(-\frac{d}{dr}+ eW + \frac{\varepsilon_{\theta}}{r}\right)\Phi _{+} -\left(m+S-eV + \varepsilon \right)\Phi _{-}} = {0}
 \end{eqnarray}
which can be written as follows
\begin{eqnarray} \label{eqq37}
 {\Phi _{+}} & {=} & {-\frac{1}{\left(m+S+ eV-\varepsilon\right) }\left(\frac{d}{dr}+eW+\frac{\varepsilon_{\theta}}{r}\right)\Phi _{-} } \label{eqq37}\\ {\Phi _{-}} & {=} & {\frac{1}{\left(m+S-eV+\varepsilon\right)}\left(-\frac{d}{dr}+eW+\frac{\varepsilon_{\theta}}{r}\right)\Phi _{+}}. \label{eqq377}
 \end{eqnarray}

Now let us consider a situation where the vector $V(r)$, pseud-scalar $W(r)$ and scalar $S(r)$ couplings
all are constant and represented by square wells or barriers. That is we have 
\begin{equation}\label{eq30}
    V(r)=\left\{\begin{array}{lll} {V_{0}} && {r\leq a} \\ {0} && {r>a} \end{array}\right., \qquad
    W(r)=\left\{\begin{array}{lll} {W_{0}} && {r\leq b} \\ {0} && {r> b} \end{array}\right., \qquad
    S(r)=\left\{\begin{array}{lll} {S_{0}} && {r\leq c} \\ {0} && {r> c} \end{array}\right.
\end{equation}
where $V_{0}$, $W_{0}$ and $S_{0}$ are constant and can be $\gtrless 0$ but $a$, $b$, $c$ are represented by positive numbers, i.e. $0<a<b<c$.
In this case our potential will be constant in each region and hence $\frac{dS}{dr}=\frac{dV}{dr}=\frac{dW}{dr}=0$ 
so that our effective equation for the lower spinor component becomes
\begin{equation}\label{30}
 \left[\frac{d^{2}}{dr^{2}}-\frac{\varepsilon_{\theta}^2+\varepsilon_{\theta}}{r^{2}}-\frac{2eW\varepsilon_{\theta}}{r}
 +\left(eV-\varepsilon\right)^{2}-\left(m+S\right)^{2}-e^2W^2 \right]\Phi _{-} =0.
\end{equation}
Similarly we obtain for the upper spinor component $\Phi _{+}$
\begin{equation}\label{31}
 \left[\frac{d^{2}}{dr^{2}}-\frac{\varepsilon_{\theta}^2-\varepsilon_{\theta}}{r^{2}}-\frac{2eW\varepsilon_{\theta}}{r}+\left(eV-\varepsilon\right)^{2}-\left(m+S\right)^{2}-e^2W^2
 \right]\Phi _{+}=0.
\end{equation}
We can combine (\ref{30}-\ref{31}) to obtain a compact form
\begin{equation}\label{eq33}
 \left[\frac{d^{2}}{dr^{2}}-\frac{\varepsilon_{\theta}^2\mp \varepsilon_{\theta}}{r^{2}}-\frac{2eW\varepsilon_{\theta}}{r}+
 \left(eV-\varepsilon\right)^{2}-\left(m+S\right)^{2}-e^2W^2\right]\Phi _{\pm} =0.
\end{equation}
Since $V$, $S$ and $W$ are constants then we can call
{\begin{equation}\label{eq344}
 \gamma^{2}=\left(m+S\right)^{2}+e^2W^2-\left(\varepsilon-eV\right)^{2}
\end{equation}}
so that our previous equation reads
\begin{equation}\label{03444}
 \left[\frac{d^{2}}{dr^{2}}-\frac{\varepsilon_{\theta}^2\mp \varepsilon_{\theta}}{r^{2}}-\frac{2eW\varepsilon_{\theta}}{r}-\gamma^{2}\right]\Phi _{\pm} =0.
\end{equation}

In the above results we have assumed that $\ga^2$ in \eqref{eq344}
is positive so that $\gamma^{2}=|\gamma|^{2}$ but for  $\gamma^{2}<0$
we can write $\gamma^{2}=-|\gamma|^{2}$. In  compact form, we have
\begin{equation}\nonumber
    \gamma=\left\{\begin{array}{lll} {|\gamma|} &\mbox{if}& {\gamma^{2}>0} \\ {i|\gamma|} &\mbox{if}& {\gamma^{2}<0} \end{array}\right.
\end{equation}
After making the change of variables $x = 2\gamma r$, we can write
\eqref{03444} as
\begin{equation}\label{3535}
 \left[\frac{d^{2}}{dx^{2}}-\frac{\mu_{1 {\pm}}^2-\frac{1}{4}}{x^{2}}+\frac{\nu_1}{x}-\frac{1}{4}\right]\Phi _{\pm}=0
\end{equation}
where we have set
\begin{equation} \label{eq40}
{\nu_1} = {-\frac{eW\varepsilon_{\theta}}{\gamma}}, \qquad
{\mu{_1}^{2}_\pm} = {\varepsilon_{\theta}\left(\varepsilon_{\theta} \mp 1\right)+\frac{1}{4}  }.
 \end{equation}
The general solution of \eqref{3535} takes the form
\beq \label{37}
\Phi _{\pm} (r) = A_{1\pm} M_{\nu_1,\mu_1}(2\gamma r) + B_{1\pm} W_{\nu_1,\mu_1}(2\gamma r)
\eeq
where $M_{\nu_1,\mu_1}(2\gamma r)$ and $W_{\nu_1,\mu_1}(2\gamma r)$ are the Whittaker hypergeometric functions.
In terms of confluent hypergeometric functions, these are given by \cite{Abramowitz}
\begin{eqnarray}
&& M_{{\nu_1,\mu_1}}(z)= e^{-z/2} z^{{\mu_1} +1/2} {}_1F_1(1/2 +\mu_1 - \nu_1{,1+2\mu_1,z})\\
&& W_{{\nu_1,\mu_1}}(z)= e^{-z/2}
z^{{\mu_1} +1/2} U(1/2 +\mu_1 -
\nu_1{,1+2\mu_1,z}).
\end{eqnarray}
The coefficients
 $A_{1\pm}$ and $B_{1\pm}$ are two constants to be defined through the boundary conditions. Before any further development,
 we would like to ponder on the general solution to our original problem
\begin{equation}
\Psi (r,\theta
)=\frac{1}{\sqrt{r}}e^{i(\varepsilon_{\theta}-\frac{1}{2} \sigma
_{3})\theta }\chi_{\pm}(r,\theta).
\end{equation}
Requiring periodic boundary condition on the total wavefunction
$\Psi (r,\theta )=\Psi (r,\theta + 2 \pi )$ gives the quantization
rule \beq \varepsilon_{\theta}= \frac{k}{2}, \qquad {k
= \pm 1, \pm 3, \pm 5, \cdots.} \eeq
 Hence the most general
solution of our problem reads
 \beq \Psi (r, \theta)=\sum_{k,\pm}
\frac{1}{\sqrt{r}}e^{\frac{i}{2}(k - \sigma _{3}) \theta
}[A_{1\pm} M_{\nu_1,\mu_1}(2\gamma r) +B_{1\pm}
W_{\nu_1,\mu_1}(2\gamma r)] \eeq which are also eigenfunctions of
the total angular momentum defined by \beq J_z =L_z +
\frac{1}{2}\sigma _{3}=-i
\partial_{\te}+ \frac{1}{2}\sigma _{3} \eeq since this operator
commutes with the full Hamiltonian of the system. It is worth
mentioning that we have reached this important result through the
use of our factorization scheme rather than the usual symmetry
approach adopted in most previous work.

Next we would like to study the potential existence of bound states whose wavefunction decreases exponentially with the variable $r$. To study confinement of the potential we should consider the asymptotic behavior for large $r$, i.e. $r\rightarrow \infty$. In this case our radial equation \eqref{03444} becomes 
\begin{equation}\label{eq50}
    \left[\frac{d^{2}}{dr^{2}}+\gamma^{2}\right]\Phi _{\pm} =0 
\end{equation}
whose solution are given by
\begin{equation}\label{eq51}
    \Phi _{\pm} (r)=A_2e^{i\gamma r}+B_2e^{-i\gamma r}
\end{equation}
which will be a propagating solution if $\gamma$ is real 
and a decaying (exponentially decaying) solution if $\gamma$ is imaginary. 

\begin{table}[h]
\begin{center}
\begin{tabular}{|c|c|c|c|c|}
\hline
Region & $r<a$ & $a<r<b$ & $b<r<c$ &$c<r$  \\
\hline
$$ & $\gamma^{2}={\beta^2-\left(\varepsilon-eV_0\right)^{2}}$ & $\gamma_1^{2}={\beta^{2}-\varepsilon^{2}}$ &
 $\gamma_2^{2}={\left(m+S_0\right)^{2}-\varepsilon^{2}}$ &$\gamma_3^{2}={m^{2}-\varepsilon^{2}}$\\
\hline
I & $|eV_0-\varepsilon|<\beta $ & $|\varepsilon|<\beta$ & $|\varepsilon|<m+S_0$& $|\varepsilon|<m$ \\
\hline
II & $|eV_0-\varepsilon|<\beta $ & $|\varepsilon|<\beta$ & $|\varepsilon|<m+S_0$& $|\varepsilon|>m$ \\
\hline
III & $|eV_0-\varepsilon|<\beta $ & $|\varepsilon|<\beta$ & $|\varepsilon|>m+S_0$& $|\varepsilon|>m$ \\
\hline
IV & $|eV_0-\varepsilon|<\beta $ & $|\varepsilon|>\beta$ & $|\varepsilon|>m+S_0$& $|\varepsilon|>m$ \\
\hline
V & $|eV_0-\varepsilon|>\beta $ & $|\varepsilon|<\beta$ & $|\varepsilon|<m+S_0$& $|\varepsilon|<m$ \\
\hline
VI & $|eV_0-\varepsilon|>\beta $ & $|\varepsilon|<\beta$ & $|\varepsilon|<m+S_0$& $|\varepsilon|>m$ \\
\hline
VII & $|eV_0-\varepsilon|>\beta $ & $|\varepsilon|<\beta$ & $|\varepsilon|>m+S_0$& $|\varepsilon|>m$ \\
\hline
VIII & $|eV_0-\varepsilon|>\beta $ & $|\varepsilon|>\beta$ & $|\varepsilon|>m+S_0$& $|\varepsilon|>m$ \\
\hline
\end{tabular}
\end{center}
\caption{\sf{Different regions for scalar coupling $S_0>0$, with $\beta=\sqrt{\left(m+S_0\right)^{2}+e^2W_0^2}$}.}\label{tab.11}
\end{table}

\begin{figure}[H]\label{cas11}
\centering
  \includegraphics[width=10cm, height=6.5cm]{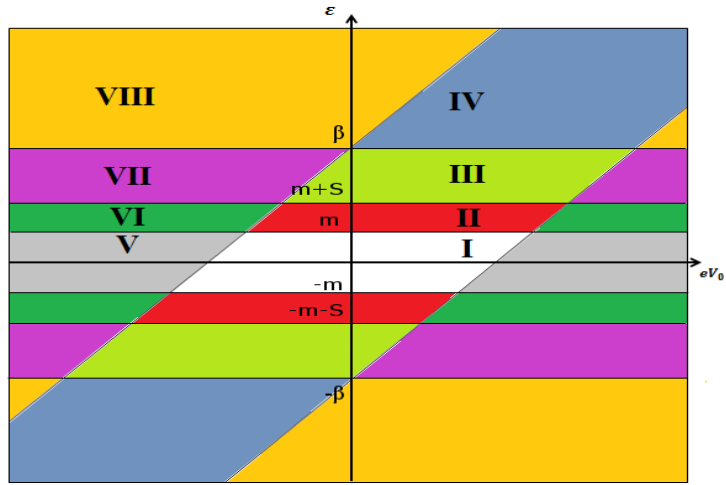}
  \caption{\sf {Representation of different region, where $S_0>0$.}}
\end{figure}


 \begin{table}[H]
 \begin{center}
\begin{tabular}{|c|c|c|c|c|}
\hline
Region & $r<a$ & $a<r<b$ & $b<r<c$ &$c<r$  \\
\hline
$$& $\gamma^{2}={\beta'^2-\left(\varepsilon-eV_0\right)^{2}}$ & $\gamma_1^{2}={\beta'^{2}-\varepsilon^{2}}$ &
 $\gamma_2^{2}={\left(m-|S_0|\right)^{2}-\varepsilon^{2}}$ &$\gamma_3^{2}={m^{2}-\varepsilon^{2}}$\\
\hline
I & $|eV_0-\varepsilon|<\beta' $ & $|\varepsilon|<\beta'$ & $|\varepsilon|<m-|S_0|$& $|\varepsilon|<m$ \\
\hline
II & $|eV_0-\varepsilon|<\beta' $ & $|\varepsilon|<\beta'$ & $|\varepsilon|<m-|S_0|$& $|\varepsilon|<m$ \\
\hline
III & $|eV_0-\varepsilon|<\beta' $ & $|\varepsilon|<\beta'$ & $|\varepsilon|>m-|S_0|$& $|\varepsilon|>m$ \\
\hline
IV & $|eV_0-\varepsilon|<\beta' $ & $|\varepsilon|>\beta'$ & $|\varepsilon|>m-|S_0|$& $|\varepsilon|>m$ \\
\hline
 V & $|eV_0-\varepsilon|>\beta' $ & $|\varepsilon|<\beta'$ & $|\varepsilon|<m-|S_0|$& $|\varepsilon|<m$ \\
\hline
 VI & $|eV_0-\varepsilon|>\beta' $ & $|\varepsilon|<\beta'$ & $|\varepsilon|>m-|S_0|$& $|\varepsilon|<m$ \\
\hline
 VII & $|eV_0-\varepsilon|>\beta' $ & $|\varepsilon|<\beta'$ & $|\varepsilon|>m-|S_0|$& $|\varepsilon|>m$ \\
\hline
VIII & $|eV_0-\varepsilon|>\beta' $ & $|\varepsilon|>\beta'$ & $|\varepsilon|>m-|S_0|$& $|\varepsilon|>m$ \\
\hline
\end{tabular}
\end{center}
\caption{\sf{Different regions for $S_0<0$ and $m>|S_0|$,  with $\beta'=\sqrt{\left(m-|S_0|\right)^{2}+e^2W_0^2}$}.}\label{tab121}
\end{table}

 \begin{table}[H]
 \begin{center}
\begin{tabular}{|c|c|c|c|c|}
\hline
Region & $r<a$ & $a<r<b$ & $b<r<c$ &$c<r$  \\
\hline
$$ & $\gamma^{2}={\beta'^2-\left(eV_0-\varepsilon\right)^{2}}$ & $\gamma_1^{2}={\beta'^{2}-\varepsilon^{2}}$ &
 $\gamma_2^{2}={\left(m-|S_0|\right)^{2}-\varepsilon^{2}}$ &$\gamma_3^{2}={m^{2}-\varepsilon^{2}}$\\
\hline
I & $|eV_0-\varepsilon|<\beta' $ & $|\varepsilon|<\beta'$ & $|\varepsilon|<|S_0|-m$& $|\varepsilon|<m$ \\
\hline
II & $|eV_0-\varepsilon|<\beta' $ & $|\varepsilon|<\beta'$ & $|\varepsilon|<|S_0|-m$& $|\varepsilon|>m$ \\
\hline
III & $|eV_0-\varepsilon|<\beta' $ & $|\varepsilon|<\beta'$ & $|\varepsilon|>|S_0|-m$& $|\varepsilon|>m$ \\
\hline
IV & $|eV_0-\varepsilon|<\beta' $ & $|\varepsilon|>\beta'$ & $|\varepsilon|>|S_0|-m$& $|\varepsilon|>m$ \\
\hline
V & $|eV_0-\varepsilon|>\beta' $ & $|\varepsilon|<\beta'$ & $|\varepsilon|<|S_0|-m$& $|\varepsilon|<m$ \\
\hline
VI & $|eV_0-\varepsilon|>\beta' $ & $|\varepsilon|<\beta'$ & $|\varepsilon|<|S_0|-m$& $|\varepsilon|>m$ \\
\hline
VII & $|eV_0-\varepsilon|>\beta' $ & $|\varepsilon|<\beta'$ & $|\varepsilon|>|S_0|-m$& $|\varepsilon|>m$ \\
\hline
VIII & $|eV_0-\varepsilon|>\beta' $ & $|\varepsilon|>\beta'$ & $|\varepsilon|>|S_0|-m$& $|\varepsilon|>m$ \\
\hline
\end{tabular}
\end{center}
\caption{\sf{Different regions for $S_0<0$ and $2m<|S_0|$, with $\beta'=\sqrt{\left(m-|S_0|\right)^{2}+e^2W_0^2}$}.}\label{tab122}
\end{table}
\begin{figure}[H]
\centering
  \includegraphics[width=8cm, height=7cm]{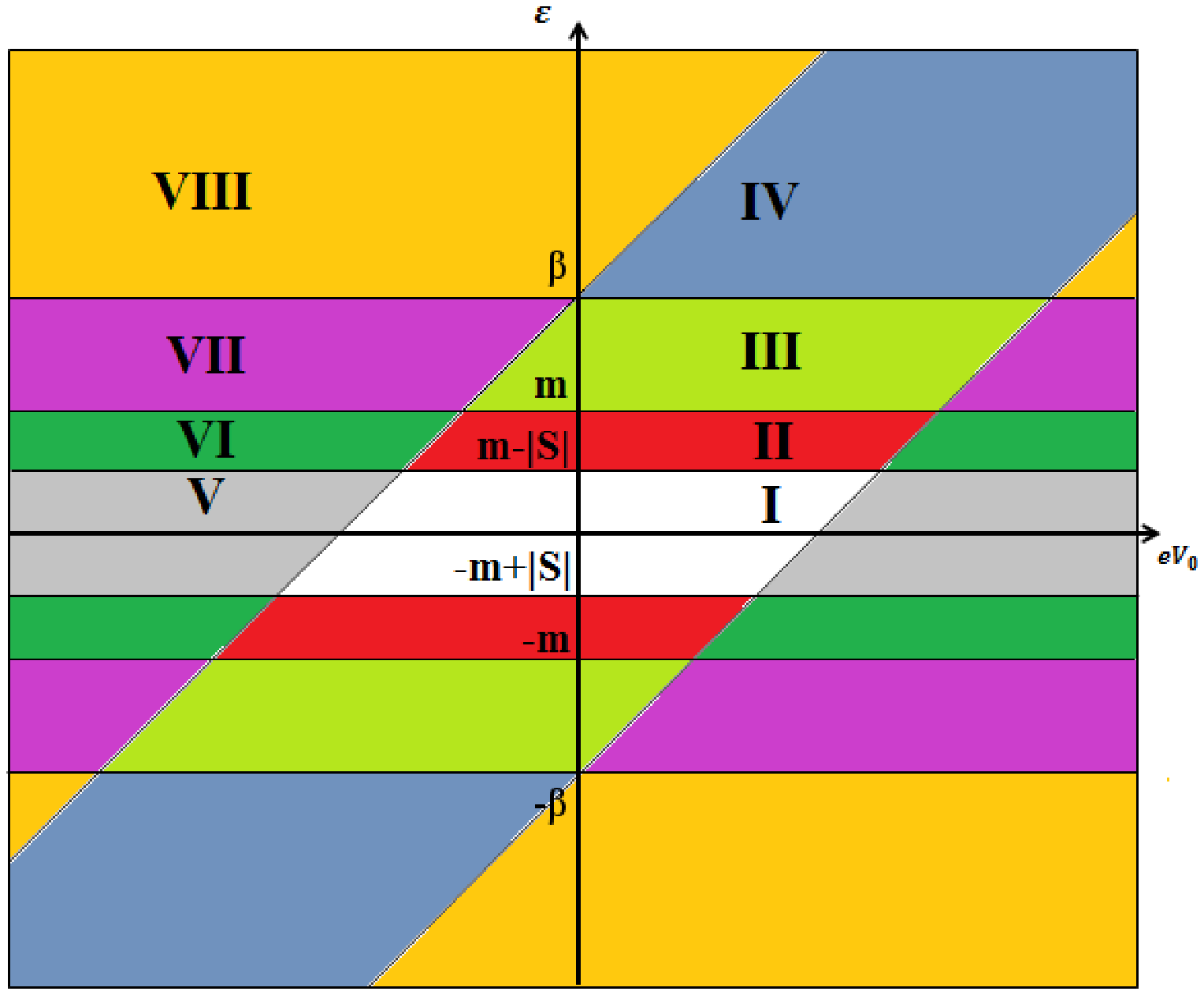}\ \ \ \ \ \
  \includegraphics[width=8cm, height=7cm]{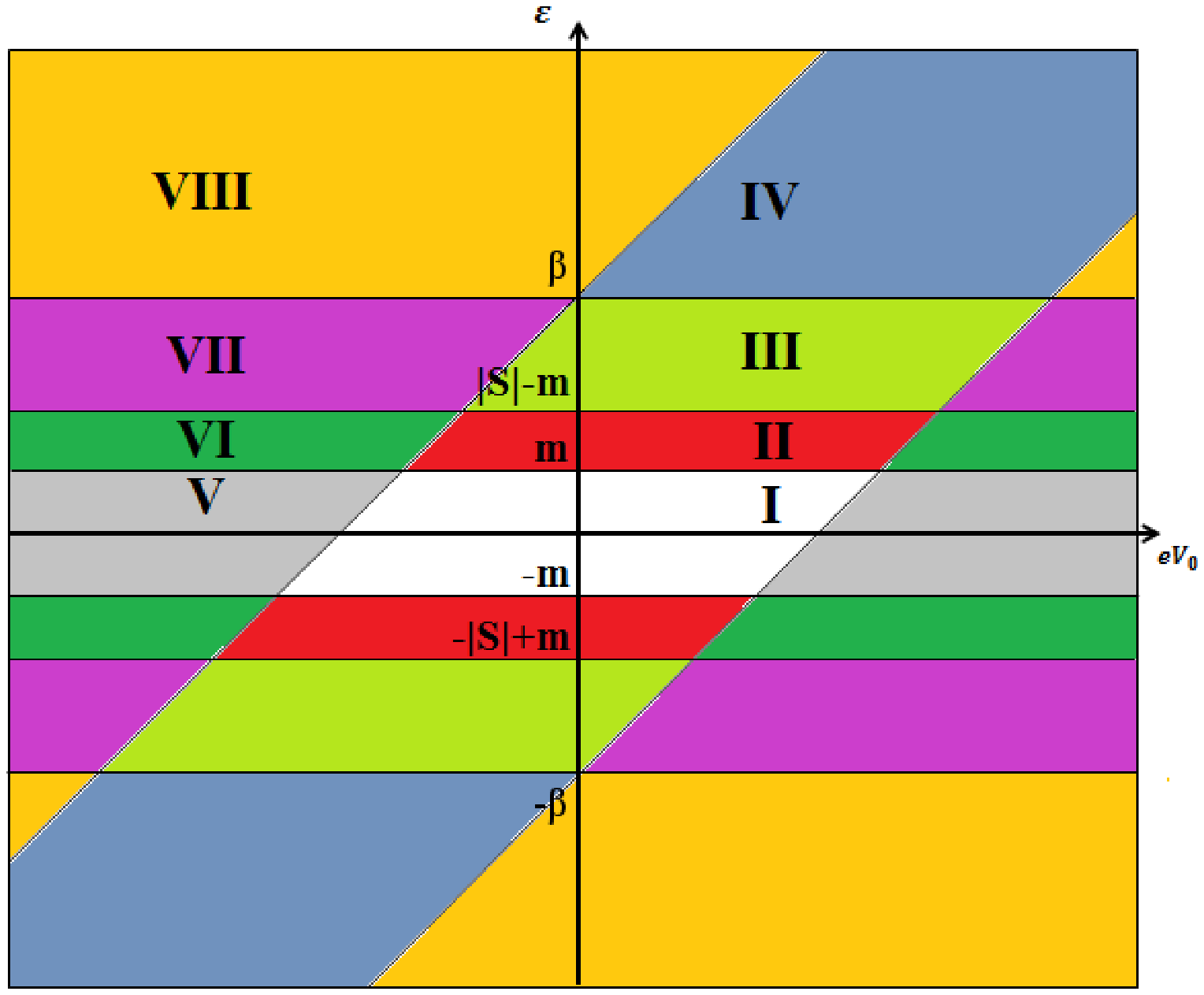}\\
  \caption{\sf {Representation of different regions for $S_0<0$ and $m>|S_0|$/$2m<|S_0|$.}}
\end{figure}
\noindent In Table \ref{tab.11}
(respectively Table \ref{tab121} and \ref{tab122})
we indicate  different regions, the decay constants are denoted by $\gamma$ (for $r < a$), $\gamma_1$ (for $a<r<b$), $\gamma_2$ (for  $b<r<c$), and $\gamma_3$ (for $r>c$)  which are imaginary or real in the cases where $S_0>0$ (respectively $S_0 < 0$, $ m > S_0$ and $S_0 < 0$, $2m < S_0$).
In the energy versus potential diagram, as shown in Figure 1, (respectively Figure 2)
one distinguishes between eight different regions. The diagram is symmetric under the transformations $\varepsilon \rightarrow -\varepsilon$ and $eV \rightarrow -eV$.

At this stage, let us examine two interesting particular cases. Indeed, the first one is the spin symmetric case $eV = S$
 and therefore \eqref{eq17}
 reduces to
  \begin{equation}
   \left(\begin{array}{cc} {m+2S-\varepsilon } & {\partial _{r}+eW+\frac{\varepsilon _{\theta } }{r} } \\ {-\partial _{r}+eW+\frac{\varepsilon _{\theta } }{r} } & {-m-\varepsilon } \end{array}\right)\left(\begin{array}{c} {\Phi _{+}} \\ {\Phi _{-} } \end{array}\right)=0 \end{equation}
giving two coupled equations
 \begin{equation}
  \begin{array}{lll} {\Phi _{+} } & {=} & {\frac{1}{m+2S-\varepsilon} \left(-\partial _{r}-eW-\frac{\varepsilon _{\theta }}{r} \right) \Phi _{-} } \\
   {\Phi _{-}} & {=} & {\frac{1}{m+\varepsilon }\left(-\partial _{r}+eW+\frac{\varepsilon _{\theta }}{r} \right) \Phi _{+} }. \end{array}
  \end{equation}
 These can be worked out to give a second order differential equation for each spinor component
 \begin{equation}
\left[\frac{\partial ^{2} }{\partial r^{2} } -\frac{\varepsilon
_{\theta }^{2} \pm \varepsilon _{\theta }}{r^{2} }-\frac{2eW\varepsilon _{\theta }}{r}
+\left(\varepsilon -m-2S\right) \left(m+\varepsilon\right)-e^2W^2\right]\Phi _{\pm } =0
\end{equation}
which can be written in compact form
\begin{equation}\label{0344}
 \left[\frac{d^{2}}{dr^{2}}-\frac{\varepsilon_{\theta}^2\mp \varepsilon_{\theta}}{r^{2}}-\frac{2eW\varepsilon_{\theta}}{r}-\eta^{2}\right]\Phi _{\pm}=0
\end{equation}
and the positive parameter ($\eta^{2}=|\eta|^{2}$) is given by
\begin{equation}
 { \eta^{2}=e^2W^2-\left(\varepsilon -m-2S\right) \left(m+\varepsilon\right)}.
\end{equation}
For $\eta^{2}<0$
we can write $\eta^{2}=-|\eta|^{2}$ so that  \eqref{0344} becomes
\begin{equation}\label{eq35}
 \left[\frac{d^{2}}{dr^{2}}-\frac{\varepsilon_{\theta}^2\mp \varepsilon_{\theta}}{r^{2}}-\frac{2eW\varepsilon_{\theta}}{r}+|\eta|^{2}\right]\Phi _{\pm} =0.
\end{equation}
This equation can be written in the following closed form
\begin{equation}\label{35}
 \left[\frac{d^{2}}{dr^{2}}-\frac{\mu_{2 {\pm}}^2-\frac{1}{4}}{r^{2}}+\frac{\nu_2}{r}-\frac{1}{4}\right]\Phi _{\pm}=0
\end{equation}
where
\begin{equation} \label{eq40}
{\nu_2} = {-\frac{eW\varepsilon_{\theta}}{\eta}}, \qquad
{\mu^{2}_{2 \pm}} = {\varepsilon_{\theta} \left(\varepsilon_{\theta} \mp 1\right) +\frac{1}{4} }.
 \end{equation}
The corresponding solution reads in terms of the Whittaker hypergeometric functions
as 
\beq \label{37}
\Phi _{\pm} (r) = A_{3\pm} M_{\nu_2,\mu_2}(2\eta r) +B_{3\pm} W_{\nu_2,\mu_2}(2\eta r)
\eeq
where
\begin{equation}\nonumber
    \eta=\left\{\begin{array}{lll} {|\eta|} &\mbox{if}& {\eta^{2}>0} \\ {i|\eta|} &\mbox{if}& {\eta^{2}<0} \end{array}\right.
\end{equation}
and in general we can write
\beq \Psi (r, \theta)=\sum_{k,\pm}
\frac{1}{\sqrt{r}}e^{\frac{i}{2}(k - \sigma _{3})\theta }[A_{3\pm}
M_{\nu_2,\mu_2}(2\eta r) +B_{3\pm} W_{\nu_2,\mu_2}(2\eta r)]. \eeq

Now we consider the second particular case, which is the pseudo-spin symmetric one such that $eV = -S$. Dirac equation  
 \eqref{eq17} reduces to 
  \begin{equation}
   \left(\begin{array}{cc} {m-\varepsilon } & {\partial _{r}+eW+\frac{\varepsilon _{\theta } }{r} } \\ {-\partial _{r}+eW+\frac{\varepsilon _{\theta } }{r} } & {-m-2S-\varepsilon } \end{array}\right)\left(\begin{array}{c} {\Phi _{+} } \\ {\Phi _{-} } \end{array}\right)=0 \end{equation}
or equivalently
 \begin{equation}
  \begin{array}{lll} {\Phi _{+} } & {=} & {\frac{1}{m-\varepsilon} \left(-\partial _{r}-eW-\frac{\varepsilon _{\theta }}{r}\right) \Phi _{-} } \\
   {\Phi _{-}} & {=} & {\frac{1}{m+2S+\varepsilon }\left(-\partial _{r}+eW+\frac{\varepsilon _{\theta }}{r}\right) \Phi _{+} } \end{array}
  \end{equation}
leading to
 \begin{equation}
\left[\frac{\partial ^{2} }{\partial r^{2} } -\frac{\varepsilon
_{\theta }^{2} \pm \varepsilon _{\theta }}{r^{2} }-\frac{2eW\varepsilon _{\theta }}{r}
+\left(\varepsilon +m+2S\right) \left(\varepsilon-m\right)-e^2W^2\right]\Phi _{\pm } =0.
\end{equation}
Similarly, we can write
\begin{equation}\label{034}
 \left[\frac{d^{2}}{dr^{2}}-\frac{\varepsilon_{\theta}^2\mp \varepsilon_{\theta}}{r^{2}}-\frac{2eW(r)\varepsilon_{\theta}}{r}-\eta'^{2}\right]\Phi _{\pm}=0
\end{equation}
by defining a positive parameter
\begin{equation}\nonumber
  {\eta'^{2}=e^2W^2-\left(\varepsilon +m+2S\right)
  \left(\varepsilon-m\right)}
\end{equation}
such as $\eta'^{2}=|\eta'|^{2}$. For $\eta'^{2}<0$
we can write $\eta'^{2}=-|\eta'|^{2}$ and \eqref{034} becomes
\begin{equation}\label{eq35}
 \left[\frac{d^{2}}{dr^{2}}-\frac{\varepsilon_{\theta}^2\mp \varepsilon_{\theta}}{r^{2}}-\frac{2eW\varepsilon_{\theta}}{r}+|\eta'|^{2}\right]\Phi _{\pm}=0
\end{equation}
which can be rearranged as follows
\begin{equation}\label{35}
 \left[\frac{d^{2}}{dr^{2}}-\frac{\mu_{3 \pm}^2-\frac{1}{4}}{r^{2}}+\frac{\nu_3}{r}-\frac{1}{4}\right]\Phi _{\pm} =0
\end{equation}
where
\begin{equation} \label{eq40}
{\nu_3} {=}  {-\frac{eW\varepsilon_{\theta}}{\eta'}}, \qquad
{\mu^{2}_{3 }\pm}  {=}  {\varepsilon_{\theta} \left(\varepsilon_{\theta} \mp 1\right) +\frac{1}{4} }.
 \end{equation}
Its solution reads
\beq \label{37}
\Phi _{\pm} (r)=A_{4\pm} M_{\nu_3,\mu_3}(2\eta' r) +B_{4\pm} W_{\nu_3,\mu_3}(2\eta' r)
\eeq
where
\begin{equation}\nonumber
    \eta'=\left\{\begin{array}{lll} {|\eta'|} &\mbox{if}& {\eta'^{2}>0} \\ {i|\eta'|} &\mbox{if}& {\eta'^{2}<0} \end{array}\right.
\end{equation}
The most general solution reads as follows \beq \Psi (r,
\theta)=\sum_{k,\pm} \frac{1}{\sqrt{r}}e^{\frac{i}{2}(k - \sigma
_{3})\theta } [A_{4\pm} M_{\nu_3,\mu_3}(2\eta' r) +B_{4\pm}
W_{\nu_3,\mu_3}(2\eta' r)]. \eeq Thus in summary we have obtained
the most general solution of the Dirac equation associated with
the first potential configuration corresponding to constant
potentials within the dot region. For bound states the solutions
are required to be finite within the dot region while they should
decay exponentially outside the dot. Continuity of the spinor
component at the dot boundary will result in the associated energy
spectrum for bound states. Scattering states, on the other hand,
require the wavefunction to have an oscillatory behavior. Before
closing this section it is worth mentioning that the Coulomb
potential for a charged impurity of electric charge $Z$ in a
uniform magnetic field along the $z$-direction pertains to this
potential configuration since it corresponds to a four vector
potential of the form
\beq
A_{0}(\vec{r})=V(r)=-\frac{Z}{r}, \qquad  A_{r}(\vec{r})=R(r)=0, \qquad A_{\theta}=W(r)=\frac{1}{2}Br. \label{eq14}
\eeq
Under these circumstances the asymptotic behavior of the radial wavefunction is monitored by the asymptotic radail equation which for
large value of $r$ reads
\begin{equation}
 \left[\frac{d^{2}}{dr^{2}}-\frac{\lambda}{r^{2}} \right]\Phi _{\pm}(r)=0, \qquad
 \lambda = \sqrt{\varepsilon_{\theta}^2 - (Ze)^2}.
\end{equation}
Hence it reduces to an inverse square potential which has very peculiar properties due to its scale invariance and lack of ground state. In fact it has been shown in this context that there is a critical value of the potential strength $\lambda_c = -1/4$ above which there are bound states and below which they do not exist. This can be realted to the solution of the above equation which is related to Bessel functions whose index is pure imaginary above the critical value and pure real below the critical value. Since the lowest value of the azimutal quatum number $\varepsilon_{\theta}= 1/2$ then this gives rise to a critical charge $K_c =1/2$, recall that $K$ can be related to the actual impurity charge $Z$ and dielectric contant $\kappa$ through  $K = Ze^2/v_F \kappa$ where $v_F$ is the Fermi velocity. This crititical behavior is related the singularity of the potential at the origin and that is why it is not sensitive to the magnetic field whose vector potential vanishes for small $r$.

\section{Second potential configuration} 

We consider the second potential configuration \eqref{eq18} and perform an analogous analysis to previous section. In this case
\eqref{GEQ136} can be written as
\beq
\label{eq19}
\left[
   \begin{pmatrix}
    {m+S+eV -\varepsilon} & {\partial _{r} +ieR} \\
    {-\partial _{r} -ieR} & {-m-S+eV -\varepsilon} \\
  \end{pmatrix}
  +\frac{1}{r}
    \begin{pmatrix}
                  {0} & {-i\partial _{\theta }+eW} \\
                  {-i\partial _{\theta }+eW} & {0} \\
              \end{pmatrix}
\right]
 \begin{pmatrix}
    {\Phi_{+}F_{+}} \\
    {\Phi_{-}F_{-}} \\
  \end{pmatrix}
=0
\eeq
where the structure of $\theta$-dependent spinor component is dictated by the
$-\frac{i}{r}\partial_{\theta}+ eW(\theta)$ angular operator so that we can factorize the angular part by requiring
%
\begin{equation}\label{eq20}
  F_{+}(\theta)= F_{-}(\theta)= F(\theta),  \qquad \left[-i\partial _{\theta }+eW(\theta) \right]F=\varepsilon_{\theta}F
\end{equation}
whose the solution is
\begin{equation}\label{eq20}
  F(\theta)= e^{ i\left[\varepsilon_{\theta } \theta + e \int W(\theta) d\theta\right]}.
\end{equation}
The periodicity of the total wavefunction requires that $\Psi (r,\theta )=\Psi (r,\theta + 2 \pi )$ and gives the quantized quantities  $\varepsilon_{\theta}$ which will dependent on the shape of the non-central part of the potential function $ W(\theta)$.
Hence it will not lead necessarily to integer or even half
integer values for the parameter $\varepsilon_{\theta}$, in general they will be represented by real numbers.
The radial part of the wavefunction on the other hand satisfies
\begin{equation}\label{eq210}
    \left(
      \begin{array}{cc}
        {m+S+eV -\varepsilon} & {\partial _{r} +ieR+\frac{\varepsilon_{\theta}}{r}} \\
        {-\partial _{r} -ieR+\frac{\varepsilon_{\theta}}{r}} & {-m-S+eV-\varepsilon} \\
      \end{array}
    \right)
   \left(
     \begin{array}{c}
       {\Phi_{+}} \\
       {\Phi_{-}} \\
     \end{array}
   \right)=0.
\end{equation}
This equation can be simplified by gauging away the spacial part of the vector potential, i.e $eR(r)$ term,
and reduces to 
\begin{equation}\label{eq250}
    \left(
      \begin{array}{cc}
        {m+S+eV -\varepsilon} & {\partial _{r}+\frac{\varepsilon_{\theta}}{r}} \\
        {-\partial _{r}+\frac{\varepsilon_{\theta}}{r}} & {-m-S+eV-\varepsilon} \\
      \end{array}
    \right)
   \left(
     \begin{array}{c}
       {\Phi_{+}} \\
       {\Phi_{-}} \\
     \end{array}
   \right)=0.
\end{equation}

From the above equations, it is clear that to go further, we need to specify
the nature of different potential couplings involved in the equation. If we adopt
the simple choice of the potential structures \eqref{eq30}, which basically describes
a quantum dot, then
%
after some algebra  we 
obtain the second order differential equations
\begin{equation}\label{eq33}
 \left[\frac{d^{2}}{dr^{2}}-\frac{\varepsilon_{\theta}(\varepsilon_{\theta}\mp1)}{r^{2}}+(eV-\varepsilon)^{2}-(m+S)^{2}\right]\Phi _{\pm}=0.
\end{equation}
Setting the  parameter 
\begin{equation}\label{eq34}
   \alpha^{2}=\left(eV-\varepsilon\right)^{2}-\left(m+S\right)^{2}
\end{equation}
we can write \eqref{eq33} as
\begin{equation}\label{eq35}
 \left[\frac{d^{2}}{dr^{2}}-\frac{\varepsilon_{\theta}(\varepsilon_{\theta}\mp1)}{r^{2}}+\alpha^{2}\right]\Phi _{\pm}=0.
\end{equation}
As before, we can make the change of variable
$X=|\alpha|r$ 
to reduce the equation to
\begin{equation}\label{eq36}
 \left[\frac{d^{2}}{dX^{2}}-\frac{\varepsilon_{\theta}(\varepsilon_{\theta}\mp1)}{X^{2}}+1\right]\Phi _{\pm}=0.
\end{equation}
This equation has some common features with the one associated with Bessel functions. To clarify this statement, let us
write the solution of \eqref{eq36} as $\Phi_{\pm}(X)=X^{\mu_4}F_{\nu_4}(X)$
to obtain
\begin{equation}\label{eq38}
  \left[\frac{d^{2}}{dX^{2}}+\frac{2\mu_4}{X}\frac{d}{dX}-\frac{\varepsilon_{\theta}(\varepsilon_{\theta} \mp 1)-\mu_4(\mu_4-1)}{X^{2}}+1\right]F_{\nu_4}=0.
\end{equation}
Comparing this with Bessel's equation $(J_{\nu_4}, H_{\nu_4}, Y_{\nu_4}, \cdots)$
\begin{equation}\label{eq39}
     \left[\frac{d^{2}}{dX^{2}}+\frac{1}{X}\frac{d}{dX}-\frac{\nu_{4 \pm}^{2}}{X^{2}}+1\right]J_{\nu_{4 \pm}}=0
\end{equation}
we realize that they will be similar if we choose the parameters $\mu_4$ and  $\nu_4$ to have the following values
\begin{equation} \label{eq40}
\mu_4 =\frac{1}{2}, \qquad
\nu_{4 \pm}^{2}= 
{\varepsilon_{\theta}(\varepsilon_{\theta} \mp 1)+\frac{1}{4} }.
\end{equation}
Hence, the general solution is a linear combination of the two independent Bessel functions. Because
$\nu_{4\pm}$ is not necessarily an integer, then we can use $J_{\nu_{4\pm}}$ and $J_{-\nu_{4\pm}}$ as independent solutions
to write
\begin{equation}\label{eq42}
 \Phi_{\pm}(r)=\sqrt{|\alpha|r}\left[A_{5\pm}J_{\nu_{4\pm}}(|\alpha|r)+B_{5\pm}J_{-\nu_{4\pm}}(|\alpha|r)\right].
\end{equation}
In the above results we have assumed that
$\alpha^{2}$ in \eqref{eq34}
is positive so that $\alpha^{2}=|\alpha|^{2}$. For $\alpha^{2}<0$
we can write $\alpha^{2}=-|\alpha|^{2}$ and 
\eqref{eq35} takes the form 
\begin{equation}\label{eq43}
 \left[\frac{d^{2}}{dr{2}}-\frac{\varepsilon_{\theta}(\varepsilon_{\theta}\mp1)}{r^{2}}-|\alpha|^{2}\right]\Phi _{\pm}=0.
\end{equation}
Using the change $X=|\alpha|r$ we obtain
\begin{equation}\label{eq44}
 \left[\frac{d^{2}}{dX^{2}}-\frac{\varepsilon_{\theta}(\varepsilon_{\theta}\mp1)}{X^{2}}-1\right]\Phi _{\pm}=0.
\end{equation}
A transformation similar to 
$ \Phi_{\pm}(X)=X^{\mu_4}F_{\nu_4}(X)$ gives
\begin{equation}\label{eq45}
  \left[\frac{d^{2}}{dX^{2}}+\frac{2\mu_4}{X}\frac{d}{dX}+\frac{\mu_4(\mu_4-1)-\varepsilon_{\theta}(\varepsilon_{\theta} \mp 1)}{X^{2}}-1\right]F_{\nu_4}=0.
\end{equation}
Comparing this equation with the modified Bessel equation
\begin{equation}\label{eq46}
     \left[\frac{d^{2}}{dX^{2}}+\frac{1}{X}\frac{d}{dX}-\frac{\nu_{4\pm}^{2}}{X^{2}}-1\right]I_{\nu_{4\pm}}=0
\end{equation}
we see that to identify \eqref{eq45} with \eqref{eq46} we need to select our parameters as follows
\begin{equation} \label{eq40}
{ \mu_4 =\frac{1}{2}}, \qquad
{ \nu_{4\pm}^{2}=\left(\varepsilon_{\theta}\mp \frac{1}{2}\right)^{2}}.
 \end{equation}
Since $I_{{\nu_4}}(X)= e^{-i\nu_4 \frac{\pi}{2}}
J_{\nu_4}(iX)$ then we  can wrap up together all cases,
$\alpha^{2}>0$ and $\alpha^{2}<0$, so that the general solution of
\eqref{eq35} is given by
\begin{equation}\label{eq49}
 \Phi_{\pm}(r)=\sqrt{\alpha r}\left[A_{6\pm}J_{\nu_{4\pm}}(\alpha r)+B_{6\pm}J_{-\nu_{4\pm}}(\alpha r)\right]
\end{equation}
where
\begin{equation}\nonumber
    \alpha=\left\{\begin{array}{lll} {|\alpha|} &\mbox{if}& {\alpha^{2}>0} \\ {i|\alpha|} &\mbox{if}& {\alpha^{2}<0}. \end{array}\right.
\end{equation}
The periodicity of the total wavefunction, $\Psi (r,\theta )=\Psi (r,\theta + 2 \pi )$, requires the condition
\beq
e^{ i\left[2\pi\varepsilon_{\theta }  + e \int W(\theta) d\theta\right]}= -1
\eeq
 which will lead to the quantized values of the quantum number $\varepsilon_{\theta }$ depending
on the explicit form of the angular potential $W(\theta)$.

Now we can study the potential existence of bound states whose wavefunction decreases exponentially for large $r$. To study confinement of the potential we should consider its asymptotic behavior for large $r$.
In this case our angular equation  \eqref{eq35} becomes 
\begin{equation}\label{eq50}
    \left[\frac{d^{2}}{dr^{2}}+\alpha^{2}\right]\Phi _{\pm}=0 
\end{equation}
and the solution is given by
\begin{equation}\label{eq51}
    \Phi _{\pm} (r)=A_7e^{i\alpha r}+B_7e^{-i\alpha r}
\end{equation}
which will be propagating solution if $\alpha$ is real,
    $\al^2>0$,
and a decaying (exponentially decaying) solution if $\alpha$ is imaginary,
    $\al^2<0$.\\

 \begin{table}[H]
 \begin{center}
\begin{tabular}{|c|c|c|c|}
\hline
Region & $r<a$ & $a<r<c$ & $c<r$  \\
\hline
$$ & $\alpha^{2}=\left(eV-\varepsilon\right)^{2}-\left(m+S\right)^{2}$ & $\alpha_1^{2}=\varepsilon^{2}-\left(m+S\right)^{2}$ &
 $\alpha_2^{2}=\varepsilon^{2}-m^{2}$ \\
\hline
I & $|eV_0-\varepsilon|<m+S_0 $ & $|\varepsilon|<m+S_0$ & $|\varepsilon|<m$ \\
\hline
II & $|eV_0-\varepsilon|<m+S_0 $ & $|\varepsilon|<m+S_0$ & $|\varepsilon|>m$ \\
\hline
III & $|eV_0-\varepsilon|<m+S_0 $ & $|\varepsilon|>m+S_0$ & $|\varepsilon|>m$ \\
\hline
IV & $|eV_0-\varepsilon|>m+S_0 $ & $|\varepsilon|<m+S_0$ & $|\varepsilon|<m$ \\
\hline
V & $|eV_0-\varepsilon|>m+S_0 $ & $|\varepsilon|<m+S_0$ & $|\varepsilon|>m$ \\
\hline
VI & $|eV_0-\varepsilon|>m+S_0 $ & $|\varepsilon|>m+S_0$ & $|\varepsilon|>m$ \\
\hline
\end{tabular}
\end{center}
\caption{Different region where $S_0>0$.}\label{tabl21}
\end{table}
\begin{figure}[H]
\centering
  \includegraphics[width=10cm, height=09cm]{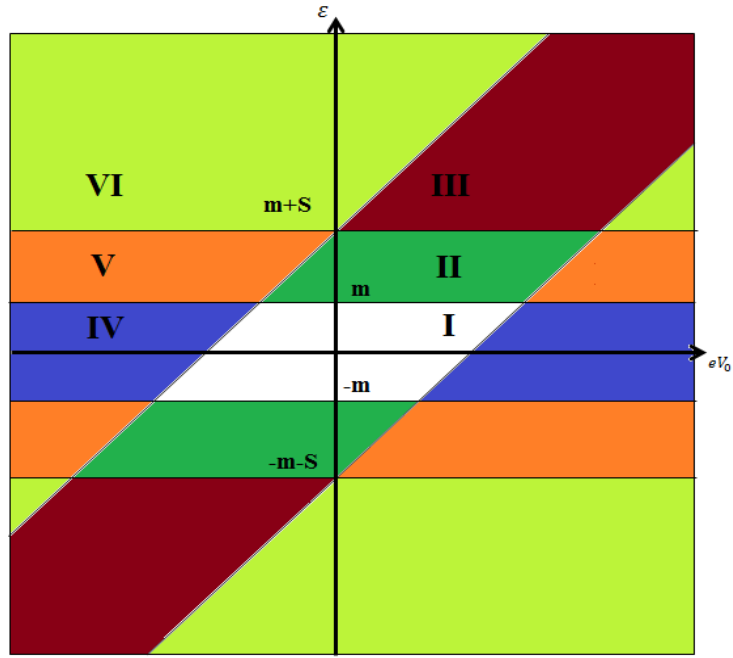}
  \caption{\sf {Representation of different cases where $S_0>0$.}}
\end{figure}
\begin{table}
 \begin{center}
\begin{tabular}{|c|c|c|c|}
\hline
Region & $r<a$ & $a<r<c$ & $c<r$  \\
\hline
$$ & $\alpha^{2}=\left(eV-\varepsilon\right)^{2}-\left(m-|S_0|\right)^{2}$ & $\alpha_1^{2}=\varepsilon^{2}-\left(m-|S_0|\right)^{2}$ &
 $\alpha_2^{2}=\varepsilon^{2}-m^{2}$ \\
\hline
I & $|eV_0-\varepsilon|<m-|S_0| $ & $|\varepsilon|<m-|S_0|$ & $|\varepsilon|<m$ \\
\hline
II & $|eV_0-\varepsilon|<m-|S_0| $ & $|\varepsilon|>m-|S_0|$ & $|\varepsilon|<m$ \\
\hline
III & $|eV_0-\varepsilon|<m-|S_0| $ & $|\varepsilon|>m-|S_0|$ & $|\varepsilon|>m$ \\
\hline
IV & $|eV_0-\varepsilon|>m-|S_0| $ & $|\varepsilon|<m-|S_0|$ & $|\varepsilon|<m$ \\
\hline
V & $|eV_0-\varepsilon|>m-|S_0| $ & $|\varepsilon|>m-|S_0|$ & $|\varepsilon|<m$ \\
\hline
 VI & $|eV_0-\varepsilon|>m-|S_0|$ & $|\varepsilon|>m-|S_0|$ & $|\varepsilon|>m$ \\
\hline
\end{tabular}
\end{center}
\caption{Different cases where $S_0<0$ and $m>|S_0|$.}\label{tabl221}
\end{table}
\begin{table}[H]
 \begin{center}
\begin{tabular}{|c|c|c|c|}
\hline
Region & $r<a$ & $a<r<c$ & $c<r$  \\
\hline
$$ & $\alpha^{2}=\left(eV-\varepsilon\right)^{2}-\left(m-|S_0|\right)^{2}$ & $\alpha_1^{2}=\varepsilon^{2}-\left(m-|S_0|\right)^{2}$ &
 $\alpha_2^{2}=\varepsilon^{2}-m^{2}$ \\
\hline
I & $|eV_0-\varepsilon|<|S_0|-m $ & $|\varepsilon|<|S_0|-m$ & $|\varepsilon|<m$ \\
\hline
 II & $|eV_0-\varepsilon|<|S_0|-m $ & $|\varepsilon|<|S_0|-m$ & $|\varepsilon|>m$ \\
\hline
III & $|eV_0-\varepsilon|<|S_0|-m $ & $|\varepsilon|>S_0|-m$ & $|\varepsilon|>m$ \\
\hline
 IV & $|eV_0-\varepsilon|>|S_0|-m $ & $|\varepsilon|<|S_0|-m$ & $|\varepsilon|<m$ \\
\hline
 V & $|eV_0-\varepsilon|>|S_0|-m $ & $|\varepsilon|<|S_0|-m$ & $|\varepsilon|>m$ \\
\hline
VI & $|eV_0-\varepsilon|>|S_0|-m$ & $|\varepsilon|>|S_0|-m$ & $|\varepsilon|>m$ \\
\hline
\end{tabular}
\end{center}
\caption{Different cases where  $S_0<0$ and $2m<|S_0|$.}\label{tabl222}
\end{table}
\begin{figure}[h]
\centering
  \includegraphics[width=8cm, height=5.5cm]{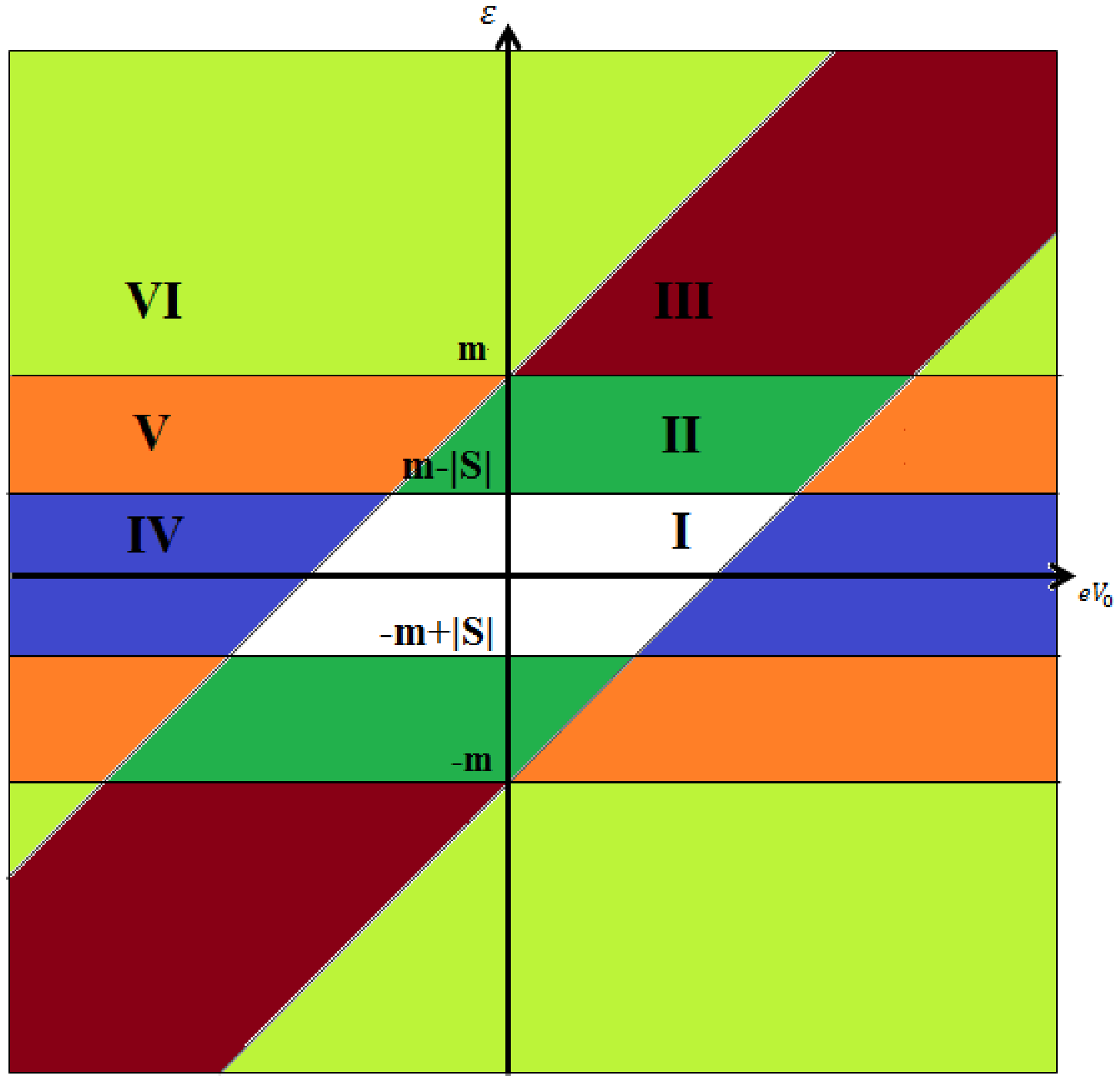}\ \ \ \ \ \
  \includegraphics[width=8cm, height=5.5cm]{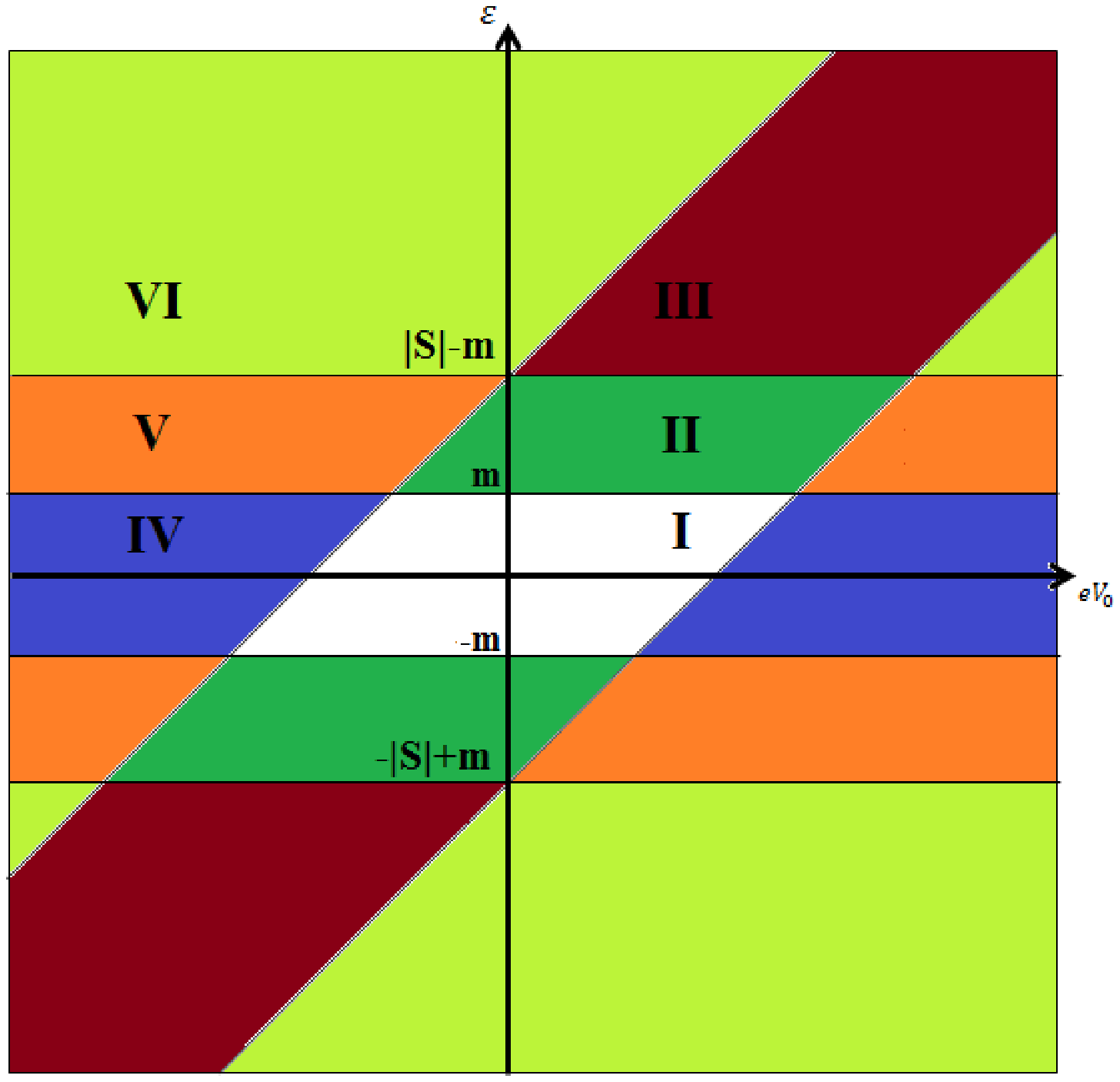}\\
  \caption{\sf {Representation of different cases where $S_0<0$, $m>|S_0|$/$2m<|S_0|$.}}
\end{figure}
\noindent In
 Table 4
 (respectively Tables 5
  and 6)
we show different regions, which are chosen according to whether $\alpha$ (for $r<a$), $\alpha_1$ (for $a<r<c$) and  $\alpha_2$ (for  $c<r$) are imaginary or real in the cases where $S_0>0$ (respectively $S_0<0$, $m>S_0$ and $S_0<0$, $2m<S_0$).
In the energy versus potential diagram, as shown in Figure 3, (respectively Figure 4)
one distinguishes between eight different regions. The diagram is symmetric under the transformations $\varepsilon \rightarrow -\varepsilon$ and $eV \rightarrow -eV$.

As before, we can also study two special cases for the present potential
configuration. The spin symmetric configuration is obtained by replacing
 $eV=S$ in \eqref{eq250} to obtain
  \begin{equation}
   \left(\begin{array}{cc} {m+2S-\varepsilon } & {\partial _{r}+\frac{\varepsilon _{\theta } }{r} } \\ {-\partial _{r}+\frac{\varepsilon _{\theta } }{r} } & {-m-\varepsilon } \end{array}\right)\left(\begin{array}{c} {\Phi _{+} } \\ {\Phi _{-} } \end{array}\right)=0 \end{equation}
which then gives
 \begin{equation}\label{eq.tho}
\left[\frac{d^{2} }{dr^{2} } -\frac{\varepsilon
_{\theta }^{2} \mp \varepsilon _{\theta }}{r^{2} }
+\tau\right]\Phi _{\pm }=0.
\end{equation}
with the parameter
\begin{equation}\label{tho}
   \tau= \left(\varepsilon -m-2S \right) \left(m+\varepsilon\right).
\end{equation}
Under the change of variable $X'=|\tau|r$
\eqref{eq.tho} becomes
 \begin{equation}\label{eq.thoo}
\left[\frac{d^{2} }{d X'^{2} } -\frac{\varepsilon
_{\theta }^{2} \mp \varepsilon _{\theta }}{X'^{2} }
+1\right]\Phi _{\pm }=0.
\end{equation}
Writing the solution as
$\Phi_{\pm}(X')=X'^{\mu_5}F_{\nu_5}(X')$
where $F_{\nu_4}(X')$ verifies
\begin{equation}\label{}
  \left[\frac{d^{2}}{dX'^{2}}+\frac{2\mu_5}{X'}\frac{d}{dX'}-\frac{\varepsilon_{\theta}(\varepsilon_{\theta} \mp 1)-\mu_5(\mu_5-1)}{X'^{2}}+1\right]F_{\nu_5}=0
\end{equation}
 and identifying  $(\mu_5,\nu_5)$
\begin{equation}
{ \mu_5 =\frac{1}{2} }, \qquad
{\nu_{5 \pm}^{2}=\left(\varepsilon_{\theta} \mp \frac{1}{2}\right)^2 }
\end{equation}
to obtain Bessel's equation
\begin{equation}\label{}
     \left[\frac{d^{2}}{dX'^{2}}+\frac{1}{X'}\frac{d}{dX'}-\frac{\nu_{5 \pm}^{2}}{X'^{2}}+1\right]J_{\nu_{5 \pm}}=0
\end{equation}
and therefore our solution can be written as
\begin{equation}
\Phi _{\pm } (r)=\sqrt{|\tau| r} \left[A_{8\pm} J_{\nu_{5\pm} }(|\tau| r)+B_{8\pm} J_{-\nu_{5\pm } }(|\tau| r)\right].
\end{equation}

For $\tau^2 <0$ we can write $\tau^2=-|\tau|^2$ and \eqref{eq250} takes the form
 \begin{equation}\label{eq.thooo}
\left[\frac{d^{2} }{d r^{2} } -\frac{\varepsilon
_{\theta }^{2} \mp \varepsilon _{\theta }}{r^{2} }
-|\tau|^2\right]\Phi _{\pm }=0.
\end{equation}
Making the change of variable $X'=|\tau|r$ and a transformation of the spinor component 
to find
\begin{equation}\label{eqbes}
  \left[\frac{d^{2}}{dX'^{2}}+\frac{2\mu_5}{X'}\frac{d}{dX'}-\frac{\varepsilon_{\theta}(\varepsilon_{\theta} \mp 1)-\mu_5(\mu_5-1)}{X^{2}}-1\right]F_{\nu_5}=0.
\end{equation}
Identifying this equation with
\begin{equation}\label{eq99}
     \left[\frac{d^{2}}{dX'^{2}}+\frac{1}{X'}\frac{d}{dX'}-\frac{\nu_{5 \pm}^{2}}{X'^{2}}-1\right]I_{\nu_{5 \pm}}=0
\end{equation}
we fix the parameters as follows 
\begin{equation}
{ \mu_5 =\frac{1}{2} }, \qquad
{\nu_{5 \pm}^{2}=\left(\varepsilon_{\theta} \mp \frac{1}{2}\right)^2 }.
\end{equation}
Since $I_{{\nu_4}}(X)= e^{-i\nu_4 \frac{\pi}{2}}
J_{\nu_4}(iX)$
the general solution 
is given by
\begin{equation}
\Phi _{\pm } (r)=\sqrt{\tau r} \left[A_{9\pm} J_{\nu_{5\pm} }(\tau r)+B_{9\pm} J_{-\nu_{5\pm } }(\tau r)\right]
\end{equation}
where
\begin{equation}\nonumber
    \tau=\left\{\begin{array}{lll} {|\tau|} &\mbox{if}& {\tau^{2}>0} \\ {i|\tau|} &\mbox{if}& {\tau^{2}<0}. \end{array}\right.
\end{equation}

Now we consider the pseudo-spin symmetric case $eV=-S$ to write \eqref{eq250} as
  \begin{equation}
   \left(\begin{array}{cc} {m-\varepsilon } & {\partial _{r}+\frac{\varepsilon _{\theta } }{r} } \\ {-\partial _{r}+\frac{\varepsilon _{\theta } }{r} } & {-m-2S-\varepsilon } \end{array}\right)\left(\begin{array}{c} {\Phi _{+} } \\ {\Phi _{-} } \end{array}\right)=0 \end{equation}
 and therefore
 \begin{equation}\label{eq109}
\left[\frac{\partial ^{2} }{\partial r^{2} } +\frac{\varepsilon
_{\theta }^{2} \pm \varepsilon _{\theta }}{r^{2} }
+\rho^2\right]\Phi _{\pm }=0.
\end{equation}
where
\begin{equation}\label{rho}
    \rho^2=\left(m+2S+\varepsilon\right)\left(\varepsilon-m\right).
\end{equation}
Under the change $X''=|\rho|r$ \eqref{eq.tho} becomes
 \begin{equation}\label{eq.109}
\left[\frac{d^{2} }{d X''^{2} } -\frac{\varepsilon
_{\theta }^{2} \mp \varepsilon _{\theta }}{X''^{2} }
+1\right]\Phi _{\pm }=0.
\end{equation}
Setting $\Phi_{\pm}(X'')=X''^{\mu_5}F_{\nu_6}(X'')$
to obtain
\begin{equation}\label{}
  \left[\frac{d^{2}}{dX''^{2}}+\frac{2\mu_6}{X''}\frac{d}{dX''}-\frac{\varepsilon_{\theta}(\varepsilon_{\theta} \mp 1)-\mu_6(\mu_6-1)}{X''^{2}}+1\right]F_{\nu_6}=0.
\end{equation}
Now if we choose $\mu_6$ and $\nu_6$ as follows
\begin{equation}
{ \mu_6 =\frac{1}{2} },\qquad
{\nu_{6 \pm}^{2}=\left(\varepsilon_{\theta} \mp \frac{1}{2}\right)^2 }
\end{equation}
we end up with
\begin{equation}\label{}
     \left[\frac{d^{2}}{dX''^{2}}+\frac{1}{X''}\frac{d}{dX''}-\frac{\nu_{6 \pm}^{2}}{X''^{2}}+1\right]J_{\nu_{6 \pm}}=0
\end{equation}
whose general solution can be written as follows
\begin{equation}
\Phi _{\pm } (r)=\sqrt{|\rho| r} \left[A_{10\pm} J_{\nu_{5\pm} }(|\rho| r)+B_{10\pm} J_{-\nu_{6\pm } }(|\rho| r)\right].
\end{equation}
For $\rho^2 <0$ we can write $\rho^2=-|\rho|^2$ and then \eqref{eq109} takes the form
 \begin{equation}\label{eq.thooo}
\left[\frac{d^{2} }{d r^{2} } -\frac{\varepsilon
_{\theta }^{2} \mp \varepsilon _{\theta }}{r^{2} }
-|\rho|^2\right]\Phi _{\pm }=0.
\end{equation}
Using the change $X''=|\rho|r$ 
to have
\begin{equation}\label{eqbes9}
  \left[\frac{d^{2}}{dX''^{2}}+\frac{2\mu_6}{X''}\frac{d}{dX''}-\frac{\varepsilon_{\theta}(\varepsilon_{\theta} \mp 1)-\mu_6(\mu_6-1)}{X^{2}}-1\right]F_{\nu_6}=0.
\end{equation}
Comparing this equation with the modified Bessel equation
\begin{equation}\label{eq999}
     \left[\frac{d^{2}}{dX''^{2}}+\frac{1}{X''}\frac{d}{dX''}-\frac{\nu_{6 \pm}^{2}}{X''^{2}}-1\right]I_{\nu_{6 \pm}}=0.
\end{equation}
Identification of \eqref{eqbes9} and \eqref{eq999} requires
\begin{equation}
{ \mu_6 =\frac{1}{2} }, \qquad
{\nu_{6 \pm}^{2}=\left(\varepsilon_{\theta} \mp \frac{1}{2}\right)^2 }.
\end{equation}
Since $I_{{\nu_4}}(X)= e^{-i\nu_6 \frac{\pi}{2}}
J_{\nu_6}(iX)$ the most general solution of \eqref{eq109} is given
by
\begin{equation}
\Phi _{\pm } (r)=\sqrt{\rho r} \left[A_{11\pm} J_{\nu_{6\pm} }(\rho r)+B_{11\pm} J_{-\nu_{6\pm } }(\rho r)\right]
\end{equation}
where
\begin{equation}\nonumber
    \rho=\left\{\begin{array}{lll} {|\rho|} &\mbox{if}& {\rho^{2}>0} \\ {i|\rho|} &\mbox{if}& {\rho ^{2}<0}. \end{array}\right.
\end{equation}

\section{Conclusion}

We have analyzed the Dirac equation in 2+1 dimensions by considering different potential couplings:
vector, pseudo-scalar and scalar.
The naive requirement of separation of variables, that is the factorization of the spinor wavefunctions
in terms of the polar coordinates $r$ and $\theta$, imposed stringent conditions of the possible structure
of potential couplings. This analysis resulted in two allowed potential configurations which were then analyzed
separately to obtain the general spinor eigenfunctions and energy spectra.

\begin{table} [H]
 \begin{center}
\begin{tabular}{|*{3}{c|}}
   \cline{1-3}
\multicolumn{3}{|c|} {First potential configuration} \\
\hline \hline
cases & solutions & parameters \\
\hline
& $A_{1\pm} M_{\nu_1,\mu_1}(2\gamma r) + B_{1\pm} W_{\nu_1,\mu_1}(2\gamma r)$ & $ \gamma^{2}=\left(eV-\varepsilon\right)^{2}-\left(m+S\right)^{2}-e^2W^2$\\
 &&$\nu_1=-\frac{eW\varepsilon_{\theta}}{\gamma}$ , \ $\mu{_1}^{2}_{\pm} =\left(\varepsilon_{\theta} \mp \frac{1}{2}\right)^2$ \\
\hline
$r\longrightarrow \infty$ &$A_2e^{i\gamma r}+B_2e^{-i\gamma r}$ & $\gamma^{2}=\left(eV-\varepsilon\right)^{2}-\left(m+S\right)^{2}-e^2W^2$\\
\hline
 $eV=S$& $A_{3\pm} M_{\nu_2,\mu_2}(2\eta r) +B_{3\pm} W_{\nu_2,\mu_2}(2\eta r)$ & $ \eta^{2}=\left(\varepsilon -m-2S\right) \left(m+\varepsilon\right)-e^2W^2$\\
&&$\nu_2=-\frac{eW\varepsilon_{\theta}}{\gamma}$ , \ $\mu{_2}^{2}_{\pm} =\left(\varepsilon_{\theta} \mp \frac{1}{2}\right)^2$  \\
\hline
$eV=-S$ & $A_{4} M_{\nu_3,\mu_3}(2\eta' r) +B_{4} W_{\nu_3,\mu_3}(2\eta' r)$ & $ \eta'^{2}=\left(\varepsilon +m+2S\right) \left(\varepsilon-m\right)-e^2W^2$\\
&&$\nu_3=-\frac{eW\varepsilon_{\theta}}{\gamma}$ , \ $\mu{_3}^{2}_{\pm} =\left(\varepsilon_{\theta} \mp \frac{1}{2}\right)^2$  \\
\hline
\end{tabular}
\end{center}
\caption{ \sf {Table summarizes the different solutions in the first potential configuration.}}\label{tabl1}
\end{table}

\begin{table} [H]
 \begin{center}
\begin{tabular}{|*{3}{c|}}
   \cline{1-3}
\multicolumn{3}{|c|} {Second potential configuration} \\
\hline \hline
cases  &solutions & parameters \\
\hline
& $\sqrt{\alpha r}\left[A_{6\pm} J_{\nu_4}(\alpha r) + B_{6\pm} J_{-\nu_6}(\alpha r)\right]$ & $ \alpha^{2}=\left(eV-\varepsilon\right)^{2}-\left(m+S\right)^{2}$\\
 &&$\nu{_6}^{2}_{\pm} =\left(\varepsilon_{\theta} \mp \frac{1}{2}\right)^2$ \\
\hline
$r\longrightarrow \infty$ &$A_7e^{i\alpha r}+B_7e^{-i\alpha r}$ & $\alpha^{2}=\left(eV-\varepsilon\right)^{2}-\left(m+S\right)^{2}$\\
\hline
 $eV=S$& $\sqrt{\tau r}\left[A_{9\pm} J_{\nu_5}(\tau r) +B_{9\pm} J_{-\nu_5}(\tau r)\right]$ & $ \tau^{2}=\left(\varepsilon -m-2S\right) \left(m+\varepsilon\right)$\\
&&$\nu{_5}^{2}_{\pm} =\left(\varepsilon_{\theta} \mp \frac{1}{2}\right)^2$  \\
\hline
$eV=-S$ & $\sqrt{\rho r}\left[A_{11\pm} J_{\nu_6}(\rho r) +B_{11\pm} J_{-\nu_6}(\rho r)\right]$ & $ \rho^{2}=\left(\varepsilon +m+2S\right) \left(\varepsilon-m\right)$\\
&&$\nu{_6}^{2}_{\pm} =\left(\varepsilon_{\theta} \mp \frac{1}{2}\right)^2$  \\
\hline
\end{tabular}
\end{center}
\caption{ \sf{Table summarizes the different solutions in the second potential configuration.}}\label{tabl1}
\end{table}
{The  tables above} summarize the nature of the radial part of the solutions for the two potential configurations
along with the spin symmetric and spin anti symmetric solutions. These solutions can studied more carefully for each
potential configuration to analyze the energy spectrum and possible existence of bound states within the dot region.

It is worth mentioning that the study of a single charged impurity embedded in a 2D Dirac equation in the presence of a uniform magnetic field was treated as a particular case of our general study. In particular we have discussed the critical behavior that is related to the singular behavior of
the potential at the origin.

\section*{Acknowledgments}

The generous support provided by the Saudi Center for Theoretical Physics (SCTP)
is highly appreciated by all authors. AJ and HB also acknowledge partial support
by  King Fahd University of Petroleum and Minerals under project
under the theoretical physics research group project RG1306-1 and RG1306-2.
AJ and HB thank the Deanship of Scientific Research at King Faisal University for funding this research number  (140232).

\section*{Appendix A: Variable separability approach to Dirac equation}

 Separability of the Dirac equation has been studied thoroughly in the past \cite{miller,kalnins,villalba}. 
 It turned out that in our context it reduces to one of the following potential forms of the spinor wavefunction:
\begin{itemize}
 \item In 2+1 dimensions:
 \beq
  \psi (r,\varphi )=\chi (\varphi )\left(\begin{array}{c} {\phi _{+} (r)} \\ {\phi _{-} (r)} \end{array}\right)\qquad \mbox{or}
  \qquad \psi (r,\varphi )=\chi (r)\left(\begin{array}{c} {\phi _{+} (\varphi )} \\ {\phi _{-} (\varphi )} \end{array}\right)\nonumber
\eeq
\item In 3+1 dimensions:
\beq
 \psi (r,\theta ,\varphi )=\chi (\varphi )\left(\begin{array}{c} {\phi _{+} (r)\left(\begin{array}{c} {\xi _{+}^{\uparrow } (\theta )} \\ {\xi _{+}^{\downarrow } (\theta )} \end{array}\right)} \\ {\phi _{-} (r)\left(\begin{array}{c} {\xi _{-}^{\uparrow } (\theta )} \\ {\xi _{-}^{\downarrow } (\theta )} \end{array}\right)} \end{array}\right)\qquad
 \mbox{or} \qquad
 \psi (r,\theta ,\varphi )=\chi (\varphi )\left(\begin{array}{c} {\phi _{+} (\theta )\left(\begin{array}{c} {\xi _{+}^{\uparrow } (r)} \\ {\xi _{+}^{\downarrow } (r)} \end{array}\right)} \\ {\phi _{-} (\theta )\left(\begin{array}{c} {\xi _{-}^{\uparrow } (r)} \\ {\xi _{-}^{\downarrow } (r)} \end{array}\right)} \end{array}\right)\nonumber
  \eeq
  \beq
\mbox{or} \qquad
 \psi (r,\theta ,\varphi )=\chi (\theta )\left(\begin{array}{c} {\phi _{+} (\varphi )\left(\begin{array}{c} {\xi _{+}^{\uparrow } (r)} \\ {\xi _{+}^{\downarrow } (r)} \end{array}\right)} \\ {\phi _{-} (\varphi )\left(\begin{array}{c} {\xi _{-}^{\uparrow } (r)} \\ {\xi _{-}^{\downarrow } (r)} \end{array}\right)} \end{array}\right) \cdots.\nonumber
 \eeq
\end{itemize}
 That is, we require that one of the variables be factorized for all $n$-spinor components, the second variable be factorized independently for the upper and lower $n/2$ - remaining spinor components and so on. The $n$-independent choice of variables will lead to different type of choices in the above factorization scheme.

\end{document}